\documentclass[11pt, epsf]{article}

\usepackage{epsfig}
\usepackage{amssymb}
\usepackage{graphicx}
\usepackage{color}
\usepackage{subfigure}

\begin{document}

\baselineskip 6mm
\renewcommand{\thefootnote}{\fnsymbol{footnote}}


\newcommand{\nc}{\newcommand}
\newcommand{\rnc}{\renewcommand}



\newcommand{\tcb}{\textcolor{blue}}
\newcommand{\tcr}{\textcolor{red}}
\newcommand{\tcg}{\textcolor{green}}


\def\be{\begin{equation}}
\def\ee{\end{equation}}
\def\ba{\begin{array}}
\def\ea{\end{array}}
\def\bea{\begin{eqnarray}}
\def\eea{\end{eqnarray}}
\def\nn{\nonumber\\}


\def\ct{\cite}
\def\la{\label}
\def\eq#1{(\ref{#1})}


\def\a{\alpha}
\def\b{\beta}
\def\g{\gamma}
\def\G{\Gamma}
\def\d{\delta}
\def\D{\Delta}
\def\e{\epsilon}
\def\et{\eta}
\def\ph{\phi}
\def\Ph{\Phi}
\def\ps{\psi}
\def\Ps{\Psi}
\def\k{\kappa}
\def\l{\lambda}
\def\L{\Lambda}
\def\m{\mu}
\def\n{\nu}
\def\th{\theta}
\def\Th{\Theta}
\def\r{\rho}
\def\s{\sigma}
\def\S{\Sigma}
\def\ta{\tau}
\def\o{\omega}
\def\O{\Omega}
\def\pr{\prime}


\def\half{\frac{1}{2}}

\def\goto{\rightarrow}

\def\na{\nabla}
\def\grad{\nabla}
\def\curl{\nabla\times}
\def\div{\nabla\cdot}
\def\pa{\partial}
\def\fr{\frac}

\def\bra{\left\langle}
\def\ket{\right\rangle}
\def\lb{\left[}
\def\lc{\left\{}
\def\ls{\left(}
\def\lp{\left.}
\def\rp{\right.}
\def\rb{\right]}
\def\rc{\right\}}
\def\rs{\right)}

\def\vac#1{\mid #1 \rangle}


\def\td#1{\tilde{#1}}
\def\check{ \maltese {\bf Check!}}


\def\Tr{{\rm Tr}\,}
\def\det{{\rm det}}


\def\bc#1{\nnindent {\bf $\bullet$ #1} \\ }
\def\ch {$<Check!>$ }
\def\ss {\vspace{1.5cm}}
\def\text#1{{\rm #1}}

\begin{titlepage}

\hfill\parbox{5cm} { }

\vspace{25mm}

\begin{center}
{\Large \bf  Holographic entanglement entropy in the nonconformal medium}

\vskip 1. cm
  {  Chanyong Park$^a$\footnote{e-mail : cyong21@sogang.ac.kr} }

\vskip 0.5cm

{\it $^a\,$ Center for Quantum Spacetime (CQUeST), Sogang University, Seoul 121-742, Korea}\\

\end{center}

\thispagestyle{empty}

\vskip2cm


\centerline{\bf ABSTRACT} \vskip 4mm

\vspace{1cm}

We investigate holographically the entanglement entropy of a nonconformal medium whose
dual geometry is described by an Einstein-Maxwell-dilaton theory. 
Due to an additional conserved charge corresponding to the number operator, its thermodynamics can be represented in a grandcanonical or canonical ensemble. 
We study thermodynamics in both ensembles by using the holographic renormalization and 
the entanglement entropy of a nonconformal medium. 
After defining the entanglement chemical potential which unlike the entanglement temperature 
has a nontrivial size dependence, we find that the entanglement entropy 
of a small subsystem satisfies the relation resembling the first law of thermodynamics in a 
medium. Furthermore,  we study the entanglement entropy change in the nonconformal
medium caused by the excitation of the ground state and by the global quench corresponding
to the insertion of particles.

\vspace{2cm}


\end{titlepage}

\renewcommand{\thefootnote}{\arabic{footnote}}
\setcounter{footnote}{0}


\section{Introduction}
 
For the last decade, there have been a huge amount of efforts to understand  strongly
interacting systems via the AdS/CFT correspondence \cite{Maldacena:1997re}. 
This new concept allowed us to study various microscopic as well as macroscopic properties 
of a conformal field theory (CFT) in the strong coupling regime, 
for example, $4$-dimensional ${\cal N}=4$ super Yang-Mills 
theory \cite{Gubser:1998bc,Witten:1998qj,Witten:1998zw} or 
$3$-dimensional ${\cal N}=6$ Chern-Simons gauge theory \cite{Bagger:2007jr,Gustavsson:2007vu}.
Its dual geometry is usually represented as an asymptotic anti de-Sitter (AdS) space with 
a proper compact manifold. These researches have been further generalized to 
the relativistic nonconformal and nonrelativistic field theory 
\cite{Kachru:2008yh}-\cite{Park:2013ana}. 
Accumulating knowledges
on those nonconformal examples would be
important in understanding the underlying structure of the gauge/gravity duality in depth
and in applying them to more realistic physical phenomena of quantum chromodynamics and 
condensed matter system. 
Moreover, in order to figure out the quantum aspect of such systems, the entanglement entropy 
becomes an important issue \cite{Ryu:2006bv}-\cite{Chakraborty:2014lfa}.
In this paper, we will study the entanglement entropy and
its thermodynamics-like behavior in a medium holographically.
  
For regarding a nonconformal theory, we should violate a scaling symmetry of
the dual geometry which can be realized by adding a scalar field called dilaton. Then,
the resulting geometry does not allow an asymptotic AdS space as a solution. Let us take into 
account an Einstein-Maxwell-dilaton theory. This gravity theory permits 
two different types of the generalized black brane solution.
One is a charged dilatonic black brane whose dual theory is mapped to a relativistic nonconformal 
theory with matter, while the other is an uncharged black brane dual to a generalized Lifshitz 
geometry where the Lifshitz scaling symmetry is broken \cite{Lee:2010qs}. 
The first corresponds to the deformed Reinssner-Nordstr\"{o}m AdS black brane 
with a nontrivial dilaton profile. The gauge field plays a different role in those two examples.
For a charged dilatonic black brane, the bulk gauge field provides an additional conserved charge representing
one of the black brane hairs, so that its dual is clearly interpreted as the number density operator
of the matter. In the generalized Lifshitz theory, the bulk field is not free and does not provide
a new black brane hair. Instead, it breaks the boost symmetry 
and generates the anisotropy between time and spatial coordinates. 
This is why the Lifshitz-type uncharged black brane 
appears \cite{Kachru:2008yh,Taylor:2008tg,Park:2013goa,Lee:2010qs}.  Anyway, since we are interested in the nonconformal medium,  
we focus on a charged dilatonic black brane from now on.

In general, it is not easy to calculate the entanglement entropy of an interacting 
quantum field theory (QFT). However, the gauge/gravity 
duality can shed light on studying the entanglement entropy even in a strong coupling regime.
In \cite{Ryu:2006bv,Ryu:2006ef,Nishioka:2009un}, it was shown that the 
holographic entanglement entropy proportional to the area of the minimal surface 
exactly reproduces the known results in a $2$-dimensional CFT \cite{Calabrese:2004eu}.
This work was further generalized to the higher dimensional cases. In an IR limit, the holographic
entanglement entropy in a black hole geometry reduces to the well-known Bekenstein-Hawking
entropy. Meanwhile, in a small subsystem corresponding to a UV limit it describes
the entanglement entropy of excited states \cite{Bhattacharya:2012mi}.
In spite of the fact that the entanglement temperature is different from
the real temperature of the system, the holographic entanglement entropy satisfies 
the thermodynamics-like relation. In a dual CFT, the entanglement temperature
is proportional to the inverse of a subsystem size, $T_E \sim 1/l$. 
This is also true for
a relativistic nonconformal QFT dual to a hyperscaling violation geometry 
\cite{Bhattacharya:2012mi}-\cite{Pang:2013lpa}. 
This fact implies that the size dependence of the entanglement temperature 
is independent of details of the theory and the entangling surface. 
From now on, we say that the entanglement temperature is universal
because it always has the same form in a relativistic dual QFT\footnote{In more general 
nonrelativistic cases with a dynamical exponent,
the size dependence of the entanglement temperature 
is further generalized to $T_E \sim 1/l^z$ \cite{Bhattacharya:2012mi}. }.

Is this still true in a medium? 
To answer this question, let us first think of thermodynamics. In a medium, 
there exists an additional conserved quantity corresponding to the number of particles.
So the first law of thermodynamics is modified into the form including the particle 
number. 
For the entanglement entropy to satisfy such a modified thermodynamic
relation, one should define a new variable representing the chemical potential
which we will call the entanglement chemical potential.
Like the entanglement temperature, it is different from the chemical potential
defined in thermodynamics. Due to the modification of the thermodynamic relation and
the new conserved charge in the medium, 
we cannot easily answer the previous question
and furthermore new issues appear. Does the entanglement entropy in a medium follow
the modified thermodynamics-like relation? If so, does the entanglement temperature
still show the same universality? Lastly, does the newly defined
entanglement chemical potential have a universal form independent of the details of the theory? 
One of goals in this paper is to clarify them.   
We find that the entanglement entropy in a medium follows the modified first thermodynamics-like
relation and that the entanglement temperature still remains universal. However, we show that the
size dependence of the entanglement chemical potential nontrivially relys on the nonconformality.
Finally, we consider the uniform insertion of particles at zero temperature which can be
regarded as a global quench deforming the original theory. 
Since this global quench modifies
the quantum states of the system, the entanglement entropy changes. 
Under such a global quench, we calculate the change of the entanglement entropy
quantitatively.

The rest of paper is organized as follows. In Sec. 2, we study
a charged black brane solution of an Einstein-Maxwell-dilaton gravity which
is dual to a nonconformal medium. From this solution, we calculate the 
thermodynamic properties in a grandcanonical and canonical ensemble.
In Sec. 3, its entanglement entropy in a small subsystem is taken into account. Due to
an additional conserved quantity, the entanglement
chemical potential is newly defined. Using it
we show that the entanglement entropy follows the first thermodynamics-like relation
and that the entanglement chemical potential has a nontrivial size dependence relying
on the nonconformality. In addition, we also investigate the change of the 
entanglement entropy under the global quench corresponding to the uniform insertion
of particles.
We finish our work with some concluding remarks in Sec. 4.

\section{Charged dilatonic black brane in the Einstein-Maxwell-dilaton theory}

Let us consider the following Einstein-Maxwell-dilaton gravity 
\cite{Lee:2010qs,Charmousis:2010zz,Charmousis:2009xr}
\begin{equation}    \la{orgact}
S= \fr{1}{2 \k^2} \int
d^{4}x\sqrt{-g}  \lb R-2(\pa \phi)^2- \fr{e^{2\alpha\phi}}{4} F_{\mu\nu}F^{\mu\nu}- 2\Lambda e^{\eta\phi}  \rb,
\end{equation}
where $\L$ denotes a negative cosmological constant.
This theory provides several different geometric solutions. If $\ph$ is a constant and 
$F_{\m\n}=0$, the simplest solution is given by a $4$-dimensional AdS space. It can 
be generalized to a Schwarzschild AdS (SAdS) black brane in the Poincare patch 
(or SAdS black hole in the global patch). The SAdS black brane is characterized by one parameter
called the black brane mass.
Turning on the gauge flux, SAdS black brane is further generalized to a
 Reissner-Nordstr\"{o}m AdS (RNAdS) black brane with two hairs, mass and charge.
 The asymptote of all these solutions is described by an AdS geometry.   
 
When $\ph$ has a nontrivial profile, the previous geometries are not solutions anymore. 
In this case, the solutions of the Einstein-Maxwell-dilaton theory are classified as follows. 
If $F_{\m\n}=0$, the Einstein-Maxwell-dilaton theory reduces to an Einstein-dilaton theory,
which allows a hyperscaling violation geometry \cite{Charmousis:2010zz}-\cite{Park:2013ana}. 
Since the overall factor of the hyperscaling violation metric breaks the scaling symmetry, 
its asymptote is not an AdS space.  
In spite of breaking of the scaling symmetry, the rotation and translation symmetries of the boundary space still survive. This fact implies that its dual field theory corresponds to a relativistic  
nonconformal QFT. In general, the hyperscaling violation geometry has a naked singularity
at the center which may indicates the instability or incompleteness of the theory. This fact
indicates that the dual  QFT is IR incomplete. To avoid this problem, one can regard
the black brane geometry. 
Analogous to a SAdS black brane, the hyperscaling violation geometry can be easily generalized 
to an uncharged black brane where the singularity is hidden behind the horizon. 
In the dual field theory at finite temperature, there is no IR incompleteness because
the Hawking temperature plays an effective IR cutoff. Even in this case, the zero temperature limit 
still remains problematic.
Another way to get rid of the IR incompleteness is to take into account a medium which
is dual of a charged black brane geometry. In this case, the dual QFT has an IR fixed point
at which the dual theory effectively becomes a $1+1$-dimensional CFT.  
As a result, a QFT with matter dual to a charged dilatonic black brane is free from
the IR incompleteness even at zero temperature.
In order to obtain a charged dilatonic black brane solution, let's turn on the gauge flux with an appropriate parameter $\a$. Then, 
we can expect that an uncharged black brane solution of the Einstein-dialton gravity  
\cite{Lee:2010qs}
is modified into a charged one
with two black brane hairs. It is true only for a specific value of $\a$. 
For general $\a$, intriguingly, there exists another uncharged black brane solution in which
the boost symmetry as well as the scaling symmetry are broken. Thus the time and spatial 
coordinate behave differently \cite{Goldstein:2009cv,Lee:2010qs}.
This is the generalization of the well-known Lifshitz geometry \cite{Kachru:2008yh}. 
In this paper, we concentrate on a QFT dua to a charged dilatonic black brane and investigate its quantum aspects described by a holographic entanglement entropy.

The equations of motion fo the Einstein-Maxwell-dilaton theory are
\begin{eqnarray}
R_{\mu\nu}-\frac{1}{2}Rg_{\mu\nu}+g_{\mu\nu} \L e^{\et \phi} &=& 2\partial_{\mu}\phi
\partial_{\nu}\phi-g_{\mu\nu}(\pa \phi)^{2}+\fr{e^{2\alpha\phi}}{2} F_{\mu\lambda}{F_{\nu}}^{\lambda}
-\frac{e^{2\alpha\phi}}{8} g_{\mu\nu} F^{2},  \la{eq:Einsteineq}\\
\fr{1}{\sqrt{-g}} \partial_{\mu}(\sqrt{-g}\partial^{\mu}\phi) &=& \frac{\et \L}{2} 
e^{\et \ph}
+\frac{\alpha \ e^{2\alpha\phi}}{8} F^{2},~~~ \\
0 &=& \partial_{\mu}(\sqrt{-g}e^{2\alpha\phi}F^{\mu\nu}) \la{eqgauge}.
\end{eqnarray}
In order to solve these equations, we take a logarithmic dilaton profile   
\be		\la{ans:dilaton}
\phi(r)= \ph_0 - \ph_1 \log r ,
\ee
where $\ph_0$ and $\ph_1$ are two integration constants.
Since $\ph_0$ can be absorbed into the cosmological constant, we can set $\ph_0 = 0$ 
without loss of generality. Now, we consider the following metric ansatz for a 
charged dilatonic black brane
\begin{equation}		\la{ans:general1}
ds^{2}=-g(r)^{2}f(r) dt^{2}+\frac{dr^{2}}{g(r)^{2} f(r)}+h(r)^{2}(dx^{2}+dy^{2}) ,
\end{equation}
with
\begin{equation}	\la{ans:general2}
g(r)=g_{0}r^{g_{1}},~~~h(r)=h_{0} r^{h_{1}} .
\end{equation}
Here the diffeomorphism allows us to set $g_0=h_0=1$. 
When we turn on a time-component gauge field $A_t$ only, 
the electric field satisfying \eq{eqgauge} is given by
\begin{equation}		\la{sol:electrifield}
F_{rt}=\frac{q}{h(r)^{2}}e^{-2\alpha\phi}.
\end{equation}
Note that a charged black brane we consider is the generalization of an uncharged black brane studied in the Einstein-dilaton theory \cite{Kulkarni:2012re}, which
preserves the boundary Lorentz symmetry. 
Since the bulk gauge field, following the gauge/gravity duality, is related to the matter, 
the dual field theory of a charged black brane corresponds to
a relativistic nonconformal QFT with the matter. In order to preserve the boundary 
Lorentz symmetry, $g_1 = h_1$ should be satisfied.
Furthermore, only when $\a = - \eta/2$, there exists
a charged black brane solution satisfying all equations of motion. In this case,
the integration constants are determined as
\be			\la{res:valinpar}
g_1 =\fr{4}{4 + \et^2} \quad , \quad  \ph_1 = \fr{2 \et}{4 + \et^2}
 \quad {\rm and} \quad \L = - \fr{4 (12 - \et^2)}{ (4 + \et^2)^2} ,
\ee
and the black brane factor is given by
\be			\la{sol:blackbrfact}
f(r)  = 1 - \frac{m}{r^{a}} + \frac{b}{r^{c}} ,
\ee
with
\bea
a =  \frac{12-\eta^2}{4+\eta^2} \quad , 
\quad b= \frac{4 + \et^2 }{16} Q^2 \quad {\rm and } \quad c= a+1= \frac{16}{4+\eta^2} ,
\eea
where $m$ and $Q$ denote two black brane hairs. 
In the $\eta=0$ limit, this charged dilatonic black brane reduces to 
an RNAdS black brane and the scaling symmetry is restored \cite{Kulkarni:2012re}. 
The near horizon geometry of the charged dilatonic black brane reduces to
$AdS_2 \times R^2$, which is independent of the nonconformality, $\et$,
and shows the existence of an IR fixed point effectively described by a $1+1$-dimensional CFT.
Above we used the nonconformality parameter to clarify the nonconformal effect, 
which is also related to the hyperscaling violation exponent \cite{Park:2013ana}
\be
\th = - \fr{2 \et^2}{4- \et^2} .
\ee

As shown in \cite{Park:2014gja}, the holographic renormalization together with 
regularity conditions of bulk fields
provide a boundary stress tensor consistent with the black brane thermodynamics. 
The regularity
of the metric requires that there is no conical singularity at the horizon and yields
the Hawking temperature 
\be			\la{res:chargedHawking}
T_H = \frac{12-\eta^2}{4  (4+\eta^2) \pi}  \ls 1 
- \frac{(4 + \et^2)^2 }{ 16 (12 - \et^2)}  \fr{Q^2}{r_h^{\fr{16 }{4 + \eta^2} }} \rs 
\ r_h^{\fr{4-\eta^2}{4+\eta^2}}  .
\ee
From the Maxwell equation, the time component of the vector field $A_t$ is determined as
\be
A_t = 2 \k^2 \m - \frac{Q}{r} ,
\ee
where $\m$ is an integration constant interpreted as a chemical potential. 
The regularity of the vector field norm at the horizon gives rise to the relation between
the chemical potential and the particle number
\be		\la{rel:Qmu}
N = 2 \k^2 V_2 \m r_h ,
\ee 
where $N = Q V_2$.

\subsection{Holographic renormalization of the grandcanonical ensemble}

Let us consider the holographic renormalization of the Einstein-Maxwell-dilaton theory, which provides
direct interpretation of the boundary energy-momentum tensor as thermodynamic quantities.
With an Euclidean signature,
the Einstein-Maxwell-dilaton action is rewritten as
\be
S_E= \int d^4 x \  {\cal L}_D ,
\ee
with
\be
{\cal L}_D =  - \fr{1}{2 \k^2}  \sqrt{g}  \lb R-2(\pa \phi)^2- \fr{e^{2\alpha\phi}}{4} F_{\mu\nu}F^{\mu\nu}- 2\Lambda e^{\eta\phi}  \rb ,
\ee
where the Euclidean metric is given by
\be
ds_E^{2}= r^{2 g_1} f(r) dt^{2}+\frac{dr^{2}}{r^{2 g_1} f(r)} + r^{2 g_1} ( dx^{2} + dy^{2} ) ,
\ee
and the Euclidean vector field  becomes
\be
A_{\ta} = - i  \ls 2 \k^2 \m - \frac{Q}{r}  \rs .
\ee
These metric and vector field together with the dilaton field in \eq{ans:dilaton} satisfy the Eucidean
equations of motion. 

In order to evaluate the on-shell gravity action, we should add several boundary terms.
The first is the Gibbons-Hawking term which is required to define the metric variation well
\be
S_{GH} = \fr{1}{\k^2} \int_{\pa {\cal M}} d^3 x \ \sqrt{\g}  \ \Th,
\ee
where $\g_{ij}$ indicates an induced metric on the boundary and the extrinsic curvature is given, 
in terms of the unit normal vector $n_{\m}$, by
\be
\Th_{\m\n} = - \half \ls \nabla_{\m} n_{\n} + \nabla_{\n} n_{\m}  \rs .
\ee
The second boundary term is a local counter term which is needed to 
make the on-shell action finite at the boundary. 
The correct counter term is
\be    \la{res:conuntt}
S_{ct} = \fr{8}{(4 + \et^2) \k^2} \int_{\pa {\cal M}} d^3 x \ \sqrt{\g} \ e^{\et \ph/2}  .
\ee
This term is the same as the one used in the holographic renormalization of the Einstein-dilaton 
theory \cite{Park:2013ana}.
Since the vector field of the charged black brane does not generate new divergence at the UV 
regime, no additional counter is required \cite{Park:2014gja}. 
If we impose a Dirichlet boundary condition on the vector field at the asymptotic
boundary, it fixes the chemical potential. In this case, all physical quantities should be 
represented as functions of the chemical 
potential and the on-shell gravity action, in the dual QFT point of view, 
is proportional to the grand potential of a grandcanonical ensemble. 
On the other hand, imposing a Neumann boundary condition instead
of a Dirichlet boundary condition is related to choose a canonical ensemble 
and requires an additional boundary term corresponding to the Legendre transformation.  

The grand potential with a Dirichlet boundary condition leads to
\bea
\O (T_H, \m, V_2) &=&  T_H \ \ls S_E + S_{GH} + S_{ct} \rs \nn
&=&  - \fr{\ls 4 - \et^2 \rs V_2}{2 \ls 4 + \et^2 \rs  \k^2} \ r_h^{(12-\et^2)/(4+\et^2)}
\ls 1 + \fr{(4+\et^2) \k^4}{4} \fr{\m^2}{r_h^{2 (4-\et^2)/(4+\et^2)}} \rs ,
\eea
where $V_2$ denotes the spatial volume at the boundary and the horizon $r_h$
becomes an implicit function of $T_H$, $\m$ and $V_2$ from \eq{res:chargedHawking} and \eq{rel:Qmu}.
The boundary energy-momentum tensor, which is obtained by varying the on-shell gravity 
action with respect
to the boundary metric
\be
{T^i}_{j} \equiv  
\lim_{r_0 \to \infty}  \ls - 2 \ \g^{i k}  \int d^3 x \  \fr{\d {\cal L}_D}{\d \g^{k j}} \rs ,
\ee
reads
\bea			\la{res:enmotensor}
E &=& {T^{0}}_{0} \nn
&=&  \fr{4  V_2}{\ls 4 + \et^2 \rs  \k^2} \ r_h^{(12-\et^2)/(4+\et^2)}
\ls 1 + \fr{(4+\et^2) \k^4}{4} \fr{\m^2}{r_h^{2 (4-\et^2)/(4+\et^2)}} \rs, \nn
P &=& - \fr{{T^{1}}_{1}}{V_2}  = - \fr{{T^{2}}_{2}}{V_2}  \nn
&=&   \fr{\ls 4 - \et^2 \rs }{2 \ls 4 + \et^2 \rs  \k^2} \ r_h^{(12-\et^2)/(4+\et^2)}
\ls 1 + \fr{(4+\et^2) \k^4}{4} \fr{\m^2}{r_h^{2 (4-\et^2)/(4+\et^2)}} \rs   .
\eea
From this result, one can see that the grand potential is related to pressure
\be
\O = - P  V_2 .
\ee

From the exact differential relation, the canonical conjugate variables of the fundamental
variables, $T_H$, $\m$ and $V_2$, are evaluated to 
\bea
S &=& - \lp \fr{\pa \O}{\pa T_H} \right|_{\m,V_2} = \fr{2 \pi V_2}{\k^2} r_h^{8/(4+\et^2)},
\la{res:thentgrand} \\
N &=& - \lp \fr{\pa \O}{\pa \m} \right|_{T_H,V_2} =  2 \k^2  V_2 r_h \m , 
\la{grand:in the ultra high energy} \\
P &=&  - \lp \fr{\pa \O}{\pa V_2} \right|_{T_H,\m} =  \fr{\ls 4 - \et^2 \rs}{2 \ls 4 + \et^2 \rs  \k^2} \ r_h^{(12-\et^2)/(4+\et^2)}
\ls 1 + \fr{(4+\et^2) \k^4}{4} \fr{\m^2}{r_h^{2 (4-\et^2)/(4+\et^2)}} \rs  .
\eea
Since the conjugate variable of temperature is the entropy, $S$ denotes
the renormalized thermal entropy derived from the renormalized grand potential. 
Intriguingly, this renormalized thermal entropy coincides with the Bekenstein-Hawking entropy. 
The particle number in \eq{grand:in the ultra high energy}
is in agreement with the regularity of the vector field. 
As a consequence, the holographic renormalization results are perfectly matched to 
those of the charged black brane thermodynamics in \eq{res:grandcanon1}.
The equation of state parameter of this system is given by
\be		\la{res:eqofstpa}
\o = \fr{P V_2}{E} = \half - \fr{\et^2}{8} .
\ee
Since the equation of parameter is independent of the chemical potential, 
it is the same as that obtained in the Einstein-dilaton theory \cite{Park:2013ana}. 
Here $\et$ indicates the nonconformality representing the deviation from the CFT.

According to the AdS/CFT correspondence, 
the conformal dimension of the dual operator for a bulk $p$-form field in $AdS_{d+1}$ 
is determined by \cite{Gubser:1998bc,Witten:1998qj}
\be
\ls \D + p \rs \ls \D + p -d \rs = m^2 ,
\ee
where the largest value of $\D$ corresponds to the conformal dimension of the dual operator.
This relation says that for $\et=0$ the dual operator of a massless bulk vector field ($p=1$ and $d=3$) has a conformal dimension $2$. In general, we may
consider many different conformal dimension $2$ operators composed of 
scalars or fermions. One example we are interested in is the operator composed of two fermions, 
${\cal O}_{\m} = \bar{\ps} \g_{\m} \ps$.
Since a fermion in a $2+1$-dimensional conformal field theory has a conformal dimension $1$,
${\cal O}_{\m}$ can be a dual operator. 
In this case, similar to the 
$5$-dimensional RNAdS black brane \cite{Park:2014gja}, we can regards the boundary value of a time-component
gauge field $A_t (z=0)$ as the chemical potential.
For a general $\et$, the relation between the bulk field and boundary operator 
is not clear. However,  if it is regarded as the nonconformal deformation from the conformal field 
theory, it may be possible to generalize the AdS/CFT correspondence to the noncoformal case.
Here, we just assume that there exists such a generalization.

The action form we consider openly appears in the string theory with specific value
of $\et$ \cite{Duff:1994an,Park:1999xn,Gubser:2009qt,Davison:2013uha}. In this case, the bulk scalar field appears as a dilaton in the string theory and
the boundary value of the dilaton field is identified with the gauge coupling
of the dual QFT. In the holographic point of view,  the nontrivial dilaton profile 
implies the nontrivial gauge coupling depending on the energy scale. For $\et=0$, since the dual theory is conformal,
the dilaton field becomes trivial. On the other hand,
\eq{res:eqofstpa} shows an explicit nonconformality for a general $\et$.  
It would be interpreted as the effect of an
irrelevant deformation or interaction because it affects on the UV behavior. 
More precisely, the deviation from the conformal theory can be read from the trace of the
energy-momentum tensor. Taking the trace of the stress tensor in \eq{res:enmotensor}, 
yields
\be
{T^i}_i = E - 2 P =   \fr{\et^2 }{\ls 4 + \et^2 \rs  \k^2} \ r_h^{(12-\et^2)/(4+\et^2)}
 + \fr{\et^2 \k^2 }{4} \ \m^2 r_h  .
\ee 
The first term shows the effect of a nonconformal interaction, 
whereas the second is the effect of the matter. These
effects disappear in the conformal limit, $\et \to 0$, as expected. 
Intriguingly,  for $\et^2=4$ the energy and pressure are reduced to
\be
E = \fr{V_2}{2 \k^2} \ls 1 + 2 \k^4 \m^2 \rs r_h \quad {\rm and} \quad P = 0 ,
\ee
which implies that the dual system is composed of pressureless particles, the so-called dust.

Now let us consider the zero temperature behavior. 
In the extremal case ($T_H=0$) except the dust, the horizon in terms of the
chemical potential is given by
\be		\la{rel:hori}
r_h = \ls \fr{(4+\et^2) \k^2 \m}{2 \sqrt{12-\et^2}}\rs^{(4+\et^2)/(4-\et^2)}.
\ee
If the dual matter is fermionic, we may regard a Fermi surface.
If there exists such a Fermi surface even in the strong coupling regime, 
the Fermi surface energy at zero temperature can be identified with the chemical
potential, $\e_F = \m$. From \eq{grand:in the ultra high energy} the Fermi surface energy
is proportional to the fermion number density, $n = N/V_2$,
\be			\la{rel:dispersionrelation}
\e_F \sim n^{(4 - \et^2)/8} .
\ee
For the conformal case, the Fermi surface energy is proportional to $\sqrt{n}$
which is similar to that of $2+1$-dimensional free fermions, although we cannot
directly compare them.

In the dust case with $\et^2 =4$, \eq{rel:hori} becomes singular so that we cannot 
apply it directly.
Instead, we should look at the metric which becomes simple for the dust
\be
g_{tt} = - \ls r - m + \fr{Q^2}{2 r} \rs .
\ee
For $m \ge \sqrt{2} Q$, it has two horizons 
\be
r_{\pm} = \fr{m \pm \sqrt{m^2 - 2 Q^2}}{2} ,
\ee
and the case saturating $m = \sqrt{2} Q$  gives rise to the extremal limit corresponding to
the zero temperature.
The Hawking temperature reads in terms of the chemical potential
\be
T_H = \fr{1}{4 \pi} \ls 1 - 2 \k^4 \m^2 \rs .
\ee
At zero temperature, the chemical potential is given by
\be			\la{rel:mukappa}
\m = \fr{1}{\sqrt{2} \k^2} .
\ee
Using this relation, the horizon is determined only by $m$
\be
r_h = \fr{m}{2} ,
\ee
where $m$ still remains as a free parameter.
The number density of the dust at zero temperature yields
\be
n =  \sqrt{2}  m  ,
\ee
where \eq{rel:mukappa} was used. This result shows that there is no direct relation between
Fermi surface energy and momentum.

\subsection{Holographic renormalization of the canonical ensemble}

As mentioned previously, if one imposes a Neumann boundary condition on the vector field, 
the charge density $Q$ is fixed. In this case, the dual system is described by a canonical ensemble.
To see this, let us vary the action with respect to the vector field.
Then, it generally generates a nontrivial boundary term, the so-called
Neumannizing term \cite{Balasubramanian:2009rx},
\be			\la{res:Neumannizing}
\d S_{E} = \fr{1}{2 \k^2} \int_{\pa {\cal M}} d^3 x \ \sqrt{g} \  e^{2\alpha\phi} g^{rr} g^{\ta \ta}  F_{r \ta} \ \d A_{\ta} .
\ee
When imposing the Dirichlet boundary condition, this term automatically vanishes
because of $\d A_{\ta}=0$. However, the Neumann boundary condition cannot get rid
of the Neumannizing term. In order to remove it, we should add a new boundary term
\be
S_{bd} = \fr{1}{2 \k^2} \int_{\pa {\cal M}} d^3 x  \ A_{\m} J^{\m} ,
\ee
with  $A_{\m} = \lc A_{\ta}, 0, 0 \rc$ and $J^{\m} = \lc i Q, 0, 0 \rc$.
The existence of this boundary term term indicates that the boundary condition changes from
Dirichlet to Neumann boundary condition
\be
\lp e^{2\alpha\phi} g^{rr} g^{\ta \ta}  F_{r \ta} \right|_{r=\infty}  = - i Q .
\ee
In the dual field theory, it is nothing but the Legendre transformation
between the grandcanonical and canonical ensembles. This fact has been crucially
used in studying the phase diagram of the holographic QCD \cite{Lee:2009bya}.

When the Neumann boundary condition is imposed, the renormalized action of a
canonical ensemble is described by
\be
S_{can} = S_E + S_{GH} + S_{ct}  + S_{bd} ,
\ee
and all quantities should be functions of $T_H$, $N$ and $V_2$.
The on-shell gravity action leads to the following the free energy 
\bea
F &=& \fr{S_{can}}{\b} \nn
&=& - \frac{\ls 4 - \eta^2 \rs V_2}{2 \ls 4 +\eta^2 \rs \k^2 }  \
r_{h}^{(12-\eta^2)/(4+\eta^2)} \ls 1 - \fr{(12+\et^2) (4 + \et^2)}{16 (4-\et^2) V_2^2}
\fr{N^2}{r_h^{16 /(4 + \eta^2)}} \rs ,
\eea
where $r_h$ is given by an implicit function of $T_H$, $N$ and $V_2$. 
The internal energy and pressure are
from the boundary energy-momentum tensor
\bea			\la{res:energywithQN}
E &=&  \fr{4 V_2}{\ls 4 + \eta ^2 \rs \k^2} \
r_h^{(12-\et^2)/(4+\et^2)}  \ls 1 + \fr{4+\et^2}{16 V_2^2} \fr{N^2}{r_h^{16/(4+\et^2)}} \rs  , \nn
P &=&
\frac{ 4 - \eta^2 }{2 \ls 4 +\eta^2 \rs \k^2 }  \
r_{h}^{(12-\eta^2)/(4+\eta^2)} \ls 1 + \fr{4 + \et^2}{16 V_2^2 }
\fr{N^2}{r_h^{16 /(4 + \eta^2)}} \rs .
\eea
Like the grandcanonical ensemble case,
all these results are in agreement with the thermodynamic results of the charged black brane in \eq{res:canon}. 
 
\section{Holographic entanglement entropy in a medium}

The entanglement entropy has been paid much attention to study quantum 
aspects of the QCD and condensed matter system. Recently, it was conjectured
following the AdS/CFT correspondence that 
the entanglement entropy of a strongly interacting system
can be understood by investigating a holographic minimal surface in the dual AdS
geometry \cite{Ryu:2006bv,Ryu:2006ef,Nishioka:2009un}. This idea on the holographic entanglement entropy is further 
generalized to non-AdS geometries dual to nonconformal field theories \cite{Nishioka:2009un,Park:2013ana}. 
In their subsequent works, intriguingly, it was shown holographically 
that the entanglement entropy in a small subsystem characterized by $l$
reveals the first thermodynamics-like relation. 
That is, the entanglement entropy of excited states follows the 
first law of thermodynamics  
\be
T_E  \  \D S_E =  \D E  ,
\ee
where $\D E$ indicates the increased energy and $T_E$ is called
the entanglement temperature. 
The entanglement temperature is different from the thermal temperature and 
is given by \cite{Bhattacharya:2012mi}
\be		\la{form:entangtemp}
T_E \sim \fr{c}{l} .
\ee
Although a constant, $c$, depends on details of the theory and
the shape of the entangling surface, the form of the entanglement temperature in \eq{form:entangtemp} is independent of them. From this point of view, 
we says that the entanglement temperature  is universal. 

This universality has been also checked in a relativistic nonconformal field theories 
without matter \cite{Park:2013ana,Bhattacharya:2012mi}. 
Unlike an uncharged black brane cases, a charged black brane has an additional
conserved quantity which usually plays an important 
role in thermodynamics. 
Depending on ensemble we choose, it becomes
a particle number or chemical potential. 
When the volume is fixed, the additional quantity 
modifies the first law of thermodynamics into
\be			\la{res:1stthemodylaw}
d E  = T  d S + \m  d N .
\ee
Similar to a thermal system, one can expect that the first thermodynamics-like law of
the entanglement entropy should also be modified in a medium. In this section, we will
investigate the entanglement entropy in a medium and check whether it satisfies the thermodynamics-like relation.

\subsection{ Holographic entanglement entropy in a strip}

Let us consider the holographic entanglement entropy of a thin strip 
\cite{Park:2013ana,Ryu:2006bv,Ryu:2006ef,Kim:2014yca}. The dual field theory of
the previous charged black brane is given by a $2+1$-dimensional relativistic 
nonconformal theory,
so we take a subsystem as a $2$-dimensional thin strip and evaluate
the entanglement entropy contained in it. 
First, we assume that the dual field theory lives in a $L^2$ spatial volume 
\be
- \fr{L}{2} \le x \le \fr{L}{2}  \quad {\rm and} \quad - \fr{L}{2} \le y \le \fr{L}{2} ,
\ee
which is a total system we consider.
Now, let us divide this system into two subsystems, $A$ and $A^c$, and 
take the area of the subsystem $A$ as a thin strip
\be			\la{par:thinstrip}
- \fr{l}{2} \le x \le \fr{l}{2}  \quad {\rm and} \quad - \fr{L}{2} \le y \le \fr{L}{2} ,
\ee
where the width of the strip, $l$, is smaller than $L$.  Following the holographic
entanglement prescription, the entanglement entropy can be evaluated
by calculating the area of the minimal surface in the dual geometry, whose boundary should 
coincide with the entangling surface of the strip. The induced metric, $h_{ij}$, on the minimal surface
becomes from \eq{ans:general1}
\be
ds^{2}=\ls \frac{r'^{2}}{r^{2 g_1} f(r)} + r^{2 g_1} \rs  dx^{2} +  r^{2 g_1}  dy^{2}  ,
\ee
where the prime indicates a derivative with respect to $x$. 
Then, the action governing the
area of the minimal surface reduces to
\bea			\la{act:minimalsurfaceact}
A 
&=& L \int_{-l/2}^{l/2} dx \ \sqrt{\fr{r'^{2}}{ f(r) }+ r^{4 g_1}} .
\eea
Due to the parity invariance under $x \to - x$, the minimal surface should have a turning point
at $x=0$ which gives rise to the minimum of $r$. If we denotes it
by $r_*$, the range spanned by the minimal surface is restricted to
$r_* \le r < r_{UV}$, where $r_{UV}$ denotes a UV cutoff.

For convenience, let us introduce dimensionless variables scaled by $r_*$
\be
z = \fr{r}{r_*} \ , \quad z_h = \fr{r_h}{r_*} \ , \quad \td{m} = \fr{m}{r_*^3}
\ , \quad  \td{b}= \fr{b}{r_*^4} \ {\rm and} \quad  z_{UV}= \fr{r_{UV}}{r_*} .
\ee
Then, the black brane factor can be rewritten as
\be
f(z) = 1 - \ls 1+ \fr{\td{b}}{z_h^{a+1}} \rs  \fr{z_h^a}{z^a}  + \fr{\td{b}}{z^{a + 1}},
\ee
where $\td{m}$ is given by a function of $\td{b}$ and $z_h$. For a small $l$,
we should take into account the case, $z_h \ll 1$. In terms of dimensionless variables, 
the Hawking temperature becomes
\be
T_H = \frac{1}{4 \pi  \ z_h \ r_*} \ls a  - \fr{\td{b}}{z_h^{a+1}} \rs .
\ee
In order to define temperature well, ${\td{b}}/{z_h^{a+1}} $ should 
be smaller than $a \sim {\cal O} (1)$. The saturation of this relation
corresponds to the extremal limit, in other words, the zero temperature limit.

The system we consider is invariant under the translation in the $x$-direction.
If regarding $x$ as a time coordinate, the Hamiltonian is conserved. 
From it, the width of the strip can be represented
as an integral form
\bea		\la{int:stripwidth}
l =  \frac{2 }{r_*^{2 g_1-1} } \int_1^{z_{UV}} d z \ \frac{1}{ z^{2 g_1} \sqrt{z^{4 g_1}-1} }
\ \frac{1}{ \sqrt{f(z)} } .
\eea 
Here the small $l$ corresponds to the large $r_*$ because $2 g_1-1 > 0$. 
Expanding $1/ \sqrt{f(z)}$ and integrating \eq{int:stripwidth} order by order, 
$r_*$ is determined as a function of $l$ 
\be
r_* =  c_0 \ l^{- \fr{1}{2 g_1 -1}}  + c_1 \ l^{\frac{a-1}{2 \text{g_1}-1}}
+ c_2 \ l^{\frac{a}{2 \text{g_1}-1}}  + {\cal O} \ls l^{\fr{a+1}{2 g_1-1 }}\rs,
\ee
where
\bea
c_0 &=& \fr{\pi ^{\frac{1}{4 \text{g_1}-2}}  \
\Gamma \left(1-\frac{1}{4 \text{g_1}}\right)^{\frac{1}{2 \text{g_1}-1}} }{2^{ \frac{1}{2 \text{g_1}-1}} \ 
g_1^{ \frac{1}{2 \text{g_1}-1}} \  \Gamma \left(\frac{3}{2}-\frac{1}{4 \text{g_1}}\right)^{\frac{1}{2 \text{g_1}-1}}}  , \nn
c_1 &=& \frac{2^{\frac{a-2 \text{g_1}}{2 \text{g_1}-1}}  \
g_1^{\frac{a-1}{2 \text{g_1}-1}}  \
\left(r_h^{a+1}+b\right) \ 
\Gamma \left(\frac{3}{2}-\frac{1}{4 \text{g_1}}\right)^{\frac{2 \text{g_1}+ a -2}{2 \text{g_1}-1}} 
\Gamma \left(\frac{a-1}{4 \text{g_1}}+1\right)}{
(2 \text{g_1}-1) \ \pi^{\frac{a-1}{4 \text{g_1}-2}}  \ \text{r_h} \
\Gamma \left(1-\frac{1}{4 \text{g_1}}\right)^{\frac{2 \text{g_1}+ a -2}{2 \text{g_1}-1}} \
\Gamma \left(\frac{6 \text{g_1}+a-1}{4 \text{g_1}}\right)} , \nn
c_2 &=& -\frac{ 2^{\frac{a-2 \text{g_1}+1}{2 \text{g_1}-1}} \ b \
g_1^{\frac{a}{2 \text{g_1}-1}} \ 
\Gamma \left(\frac{3}{2}-\frac{1}{4 \text{g_1}}\right)^{\frac{a}{\frac{2 \text{g_1}+ a -1}{2 \text{g_1}-1} }}  \  
\Gamma \left(\frac{a}{4 \text{g_1}}+1\right)}{
(2 \text{g_1}-1) \ \pi ^{ \frac{a}{4 \text{g_1}-2}} \ 
\Gamma \left(1-\frac{1}{4 \text{g_1}}\right)^{\frac{2 \text{g_1}+ a -1}{2 \text{g_1}-1}} \
\Gamma \left(\frac{a}{4 \text{g_1}}+\fr{3}{2} \right)} .
\eea
Substituting these results into the action in \eq{act:minimalsurfaceact}, the
holographic entanglement entropy reads  perturbativly 
\bea   	\la{res:holographic entanglement entropy}
S_E &\equiv& \fr{2 \pi A}{\k^2} 
= \fr{4 \pi L}{\k^2}  \  r_{UV}
+ s_0  \ l^{- \frac{1}{2 \text{g_1}-1}}
+ s_1 \ l^{\frac{a-1}{2 \text{g_1}-1}} 
+ s_2 \  l^{\frac{a}{2 \text{g_1}-1}} + \cdots    ,
\eea
where ellipsis implies higher order terms and coefficients, $s_0$, $s_1$ and $s_2$, are
given by
\bea			
s_0 &=&- \sqrt{\pi } L  \ \frac{2 \ \text{c_0} \ \Gamma \left(1-\frac{1}{4 \text{g_1}}\right) }{\Gamma \left(\frac{1}{2}-\frac{1}{4 \text{g_1}}\right)}, \nn
s_1 &=& \sqrt{\pi } L  \lc 
\frac{b \ \Gamma \left(\frac{a+4 \text{g_1}-1}{4 \text{g_1}}\right)}{(a-1) \ r_h \ 
c_0^{a-1} \  \Gamma \left(\frac{a+2 \text{g_1}-1}{4 \text{g_1}}\right)}+\frac{ r_h^a \ \Gamma \left(\frac{a+4 \text{g_1}-1}{4 \text{g_1}}\right)}{(a-1) \
c_0^{a-1}  \ \Gamma \left(\frac{a+2 \text{g_1}-1}{4 \text{g_1}}\right)}
-\frac{2 \ \text{c_1} \ \Gamma \left(1-\frac{1}{4 \text{g_1}}\right)}{\Gamma \left(\frac{1}{2}-\frac{1}{4 \text{g_1}}\right)} \rc   , \nn
s_2 &=& - \sqrt{\pi } L  \lc \frac{b \ \Gamma \left(\frac{a}{4 \text{g_1}}+1\right)}{a \
c_0^{a} \ \Gamma \left( \frac{a}{\text{4 g_1}}+ \fr{1}{2} \right)}
+\frac{2 \ \text{c_2} \ \Gamma \left(1-\frac{1}{4 \text{g_1}}\right)}{\Gamma \left(\frac{1}{2}-\frac{1}{4 \text{g_1}}\right)} \rc .
\eea
The first term in \eq{res:holographic entanglement entropy} shows the expected 
divergence of a $2+1$-dimensional field theory, which is originated from the 
short range correlation near the boundary of the strip. The remainders are related to
the long range correlation between the inside and outside of the strip.

\subsection{Conformal medium with $\et=0$}

Let us first consider the simplest conformal case.
For $\et=0$, an Einstein-Maxwell-dilaton theory reduces to an Einstein-Maxwell theory and
its geometric solution reduces to the well-known RNAdS black brane
\be
ds^2 = - r^2 f(r) dt^2 + \fr{dr^2}{r^2 f(r)} +  r^2 \ls dx^2 + dy^2 \rs ,
\ee
with
\be
f(r) = 1 - \fr{m}{r^3} + \fr{Q^2}{4 r^4}  .
\ee
When $Q=0$, it is further reduced to the SAdS black brane.
If $m$ also vanishes, the geometry becomes a pure AdS space which can be regarded as
a zero temperature limit of the SAdS black brane.
Its dual field theory is in the vacuum of a CFT.

If turning on $m$ in a small $l$ limit, the vacuum states become excited \cite{Bhattacharya:2012mi}. In this case, 
the holographic entanglement entropy of the subsystem $A$ reads
from \eq{res:holographic entanglement entropy}
\be			\la{res:confEE}
S_E = \frac{4 \pi  r_{UV}  L}{\kappa ^2}
-\frac{8 \pi ^2 \ \Gamma \left(\frac{3}{4}\right)^2 L }{\Gamma \left(\frac{1}{4}\right)^2 \kappa ^2 }  \  \fr{1}{l}+
\frac{\pi    \ \Gamma \left(\frac{1}{4}\right)^2 L r_h^3}{16  \ \Gamma \left(\frac{3}{4}\right)^2 \kappa ^2} \ l^2  ,
\ee
where $r_h  = m^{1/3}= \fr{4 \pi T_H}{3}$. The first two terms correspond to the
vacuum entanglement entropy, while
the last indicates
the increased entanglement entropy for the excited state
\be
\D S_E = \frac{\pi    \ \Gamma \left(\frac{1}{4}\right)^2 L r_h^3}{16  \ \Gamma \left(\frac{3}{4}\right)^2 \kappa ^2} \ l^2 .
 \ee 
 The energy used to excite the vacuum states of the subsystem $A$ is from \eq{res:energywithQN} with $Q=0$
\be
\D E =  \fr{ L}{\k^2} \ l \ r_h^{3}  ,
\ee
where the volume of the subsystem $A$ is given by $V_2 = l L$. Defining the entanglement temperature $T_E$ as
\be			\la{res:enttempRNAdS}
T_E = \fr{16  \ \Gamma \left(\frac{3}{4}\right)^2}{\pi  \ \Gamma \left(\frac{1}{4}\right)^2}
\ \fr{1}{l} ,
\ee
above quantities satisfy the thermodynamics-like relation
\be
\D E = T_E  \D S_E  .
\ee
As mentioned before, the entanglement temperature shows a universal behavior 
inversely proportional to the width of the strip.

Now, let us move on the medium case. In this case, since there exists an additional conserved
charge, we cannot naively compare the entanglement entropy of a medium with that of the
vacuum. In the conformal  medium with $\et=0$ and $Q \ne 0$, the 
electric charge of the black brane, $Q$, is dual to the number density of particles in the dual
CFT, 
either $Q= N / L^2$ for the total system or $Q= N_A / (l L)$ for the subsystem $A$. 
In terms of $N_A$,
the lowest order terms of the entanglement entropy are rewritten as
\be			\la{res:HEERNAdSconformal}
S_E = 
\frac{4 \pi   L}{\kappa ^2} \  r_{UV}
-\frac{8 \pi ^2 \ \Gamma \left(\frac{3}{4}\right)^2  L}{ \Gamma \left(\frac{1}{4}\right)^2 \kappa ^2 } \fr{1}{l}
+  \frac{\pi  \ \Gamma \left(\frac{1}{4}\right)^2 \lb 1+ N_A^2 /(4 l^2 L^2 r_h^4)  \rb L  r_h^3  }{16\ \Gamma \left(\frac{3}{4}\right)^2 \kappa ^2 } \  l^2
-\frac{  \Gamma \left(\frac{1}{4}\right)^2 N_A^2 }{40  \ \Gamma \left(\frac{3}{4}\right)^2 \kappa ^2 L} \  l ,
\ee
where $r_h$ is regarded as an implicit function of $T_H$ and $N_A$ following the relation
\be			\la{res:Hawkingtemprel}
T_H = \frac{3}{4 \pi }  r_h \ls 1 
- \frac{1 }{ 12 }  \fr{ N_A^2 }{l^2 L^2 \ r_h^{4}} \rs .
\ee
If one varies $T_H$ with a fixed particle number, 
it yields the entanglement entropy variation 
\bea			\la{res:entropychangefixedN}
\lp \D S_E \right|_{N_A} &=& S_E (T_H,N_A)- S_E (0,N_A) \nn
&=& \frac{\pi   \ \Gamma \left(\frac{1}{4}\right)^2 L}{
16  \ \Gamma \left(\frac{3}{4}\right)^2 \kappa ^2 }  \lb 
\ls 1 + \fr{ N_A^2}{4 l^2 L^2 r_h^4}   \rs  r_h^3
-  \ls 1 + \fr{ N_A^2}{4 l^2 L^2 r_0^4}   \rs   r_0^3 \rb  l^2 ,
\eea
where $S_E (T_H,N_A)$ and $S_E (0,N_A)$ indicate the entanglement entropy
of excited and ground states respectively
and $r_0$ implies the horizon in the extremal limit, $r_0^4 = N_A^2/(12 l^2 L^2)$.
Above only the third term in \eq{res:HEERNAdSconformal} contributes
the entanglement entropy variation through $r_h$. 
In the same system, the increased energy comes from the holographic renormalization in\eq{res:energywithQN}
\be
\lp \D E \right|_{N_A} = \fr{L}{\k^2}  \lb 
\ls 1 + \fr{ N_A^2}{4 l^2 L^2 r_h^4}   \rs  r_h^3
-  \ls 1 + \fr{ N_A^2}{4 l^2 L^2 r_0^4}   \rs   r_0^3 \rb   l  .
\ee
Intriguingly, this result shows that the increased energy is proportional to the variation of the
entanglement entropy.
When the entanglement temperature is defined as the previous one in \eq{res:enttempRNAdS}, the conformal medium satisfies the following thermodynamics-like relation 
\be			\la{res:HEEforfixedN}
\lp  \D E \right|_{N_A}  = T_E \lp  \D S_E \right|_{N_A}   .
\ee
Furthermore, the third term in \eq{res:HEERNAdSconformal}
can be reinterpreted as $\fr{E}{T_E}$.

In the medium, the additional conserved charge provides another situation. Even when
temperature of the system is not changed, the entanglement entropy can vary by adding particles. 
To see this, let us vary the particle number without changing temperature.
From the relation in \eq{res:Hawkingtemprel}, 
no variation of temperature gives rise to a relation between variations of $r_h$ and $N_A$ 
\be
\D r_h =  \frac{ N_A }{6 \lb1 +N_A^2 / (4 l^2 L^2 r_h^4 ) \rb  l^2 L^2 r_h^3} \ \D  N_A .
\ee
From this, the change of the entanglement entropy reads
\be
\lp \D S_E \right|_{T_H} =  \ls \frac{ \pi  \Gamma \left(\frac{1}{4}\right)^2 \lb 1+N_A^2 /(12 l^2 L^2 r_h^4) \rb N_A}{
16 \ \Gamma \left(\frac{3}{4}\right)^2 \lb 1+N_A^2/(4 l^2 L^2 r_h^4) \rb \kappa ^2  L  r_h} 
-\frac{\Gamma \left(\frac{1}{4}\right)^2 N_A \  l }{20 \ \Gamma \left(\frac{3}{4}\right)^2 \kappa ^2 L } \rs \D  N_A   ,
\ee
where $\lp \D S_E \right|_{T_H}$ implies the entropy change at fixed $T_H$.
and the variation of $N_A$ can be understood as a global quench, as will be shown. 
The origin of the first term is the third term in \eq{res:HEERNAdSconformal}, so it 
is related to the energy variation when varying the particle number. The last 
comes from the fourth term in \eq{res:HEERNAdSconformal}, which
is independent of the energy variation. Using the entanglement temperature, this relation
can be rewritten as
\be                \la{res:HEEforfixedTH}
T_E  \ \lp  \D S_E \right|_{T_H}  =   \lp \D E \right|_{T_H}  -   \lp  \m_E \right|_{T_H} \ \D N_A  ,
\ee
where the increased energy, $\lp \D E \right|_{T_H}$, and  the entanglement chemical potential, 
$\lp \m_E \right|_{T_H}$,  are
are given by
\bea             
\lp \D E \right|_{T_H}  &=&  \frac{  \lb 1+N_A^2 /(12 l^2 L^2 r_h^4) \rb N_A}{
3 \lb 1+N_A^2/(4 l^2 L^2 r_h^4) \rb \kappa ^2 l L  r_h}  \  \D N_A , \nn
\lp \m_E \right|_{T_H} &=& \frac{4 N_A}{5 \pi  \kappa ^2 L} .
\eea
In the conformal medium, 
the entanglement chemical potential is independent of the strip width, $l$,
and proportional to the number of particles contained in the subsystem $A$. 
Due to the fact that the entanglement temperature is universal,
it is interesting to ask whether the entanglement chemical potential also shows
a similar universal behavior in more general cases. In the next section, we will discuss on the
universality of this entanglement chemical potential in the hyperscaling violation geometry. 
Combining \eq{res:HEEforfixedN} and \eq{res:HEEforfixedTH}, the total change of the 
entanglement entropy satisfies the first law of thermodynamics in the medium 
\be		\la{res:fullthrel}
T_E \ \D S_E =  \D E -  \m_E  \ \D N _A ,
\ee
with
\bea
\D E &=& \lp  \D E \right|_{N_A}  + \lp  \D E \right|_{T_H} , \nn
\D S_E &=& \lp  \D S_E \right|_{N_A}  + \lp  \D S_E \right|_{T_H},
\eea
where $\lp \D E \right|_{N_A}$ and $\lp \D S_E \right|_{N_A}$ come from the
excitation of ground states, while the global quench gives rise to
$\lp \D E \right|_{T_H}$ and $\lp \D S_E \right|_{T_H}$.
Eq.\eq{res:fullthrel} governs the most general case including both the excitation of ground states 
and the global quench.

In order to understand the above thermodynamics-like relation at zero temperature, 
let us first suppose that the matter of a medium resides in the ground state at zero temperature. 
Following the AdS/CFT correspondence, its number density can be reinterpreted
as the dual operator of the background gauge field in the extremal RNAdS black brane geometry, 
$Q= \ps^+ \ps$. If the matter is distributed uniformly, $Q$ can be regarded
a global operator independent of the position. Then, the number of the matter in the medium
is given by $N = \int d^2 x \ Q=Q L^2$ in the total system and by $N_A = Q l L$ 
in the subsystem $A$. Now, let us deform this medium by adding $\D N_A$ particles
without exciting the ground state. This deformation corresponds to the global quench
in the medium. 
Using the regularity of the bulk gauge field $N_A \sim l L \m^{8/(4-\et^2)}$, the insertion of particles can be reinterpreted 
as the change of the chemical potential in a grandcanonical ensemble. 
For $T_H=0$, the ground state is not excited so that
there is no increment of the entanglement entropy and energy caused by the excited state,
$\lp \D S_E \right|_{N_A}=\lp \D E \right|_{N_A}=0$.
Even at zero temperature, however, the global quench can increase the system energy
and the entanglement entropy following the thermodynamics-like relation in \eq{res:HEEforfixedTH}.  
Since $r_h^4 = N_A^2/(12 l^2 L^2)$ at zero temperature, 
the increased energy and entanglement entropy change under the global quench are 
\bea
\lp \D E \right|_{T_H}   &=&  \frac{\sqrt[4]{3}  \ \sqrt{N_A}}{\sqrt{2} \ \kappa ^2 \sqrt{l} \sqrt{L}} \ \D N_A , \nn
\lp \D S_E  \right|_{T_H} &=& \frac{ \Gamma \left(\frac{1}{4}\right)^2 
\left(5  \ \sqrt[4]{12}  \ \pi \ \sqrt{l L N_A}  -8 \ l N_A \right)}{
160  \ \Gamma \left(\frac{3}{4}\right)^2 \kappa ^2 L} \ \D  N_A  .
\eea
In the small $l$ limit, the first term in the entanglement entropy change becomes dominant,
so the global quench creates the following energy and entanglement entropy 
approximately
\bea
\lp \D E  \right|_{T_H}  &\sim&  \sqrt{\fr{N_A}{l L}} \ \D  N_A , \nn
\lp \D S_E  \right|_{T_H} &\sim&  \sqrt{\fr{l N_A}{L}} \ \D  N_A .
\eea

\subsection{Nonconformal medium with a general $\et$}

Due to the universal behavior of the entanglement temperature, as mentioned before,
it would be interesting to ask whether the size dependence of the entanglement chemical potential 
is independent of the detail of the theory. In this section, we will check this point
and investigate how the global quench in a nonconformal medium changes
the entanglement entropy. This result would be useful to figure out the quantum entanglement 
of the matter states and helpful to understand the real condensed matter
system at low temperature.

From \eq{res:holographic entanglement entropy}, for a general $\et$ 
the entanglement entropy in the strip is given by
\be			\la{res:ententhsg}
S_E  
= \fr{4 \pi L }{\k^2} \  r_{UV}
+ s_0  \ l^{-\frac{4+\eta ^2}{4-\eta ^2}}
+ s_1 \ l^{2} 
+ s_2 \  l^{\frac{ 4+\eta ^2}{4-\eta ^2} } + \cdots 
\ee
with
\bea
s_0 &=& -\frac{
\pi ^{\frac{8 -\eta ^2}{4 -\eta ^2}} 
\left(4 + \eta ^2\right)^{\frac{4+\eta ^2}{4-\eta ^2}} \
\Gamma \left(\frac{3}{4}-\frac{\eta ^2}{16}\right)^{\frac{8}{4-\eta ^2}} \ L}{
2^{\frac{4+5 \eta ^2}{4-\eta ^2}}  \ 
\left(\frac{1}{4}-\frac{\eta ^2}{16}\right)^{\frac{4+\eta ^2}{4-\eta ^2}} \
\Gamma \left(\frac{1}{4}-\frac{\eta ^2}{16}\right)^{\frac{8}{4-\eta ^2}} \ \kappa ^2  } , \nn
s_1 &=& \frac{
\left(4-\eta ^2\right)^2    
 \ \Gamma \left(\frac{1}{4}-\frac{\eta ^2}{16}\right)^3 
\lb  1 + \left(4+ \eta ^2\right) N^2/ \ls 16 l^2 L^2 \ r_h^{\frac{16}{4+\eta ^2}}   \rs \rb
\ L  \ r_h^{\frac{12-\et^2}{4+\eta ^2}} }{
2^{\frac{7}{2}+\frac{\eta ^2}{8}}  \left(8-\eta ^2\right) \left(4+\eta ^2\right)  \ \Gamma \left(1-\frac{\eta ^2}{8}\right) \Gamma \left(\frac{3}{4}-\frac{\eta ^2}{16}\right) \kappa ^2 }
, \nn
s_2 &=&- \fr{
\pi ^{\frac{4- 3 \eta ^2}{2 (4-\eta ^2)}} 
\left(4-\eta ^2\right)^{\frac{8}{4-\eta ^2}} \
\Gamma \left(\frac{1}{2}-\frac{\eta ^2}{8}\right) \
\Gamma \left(\frac{1}{4}-\frac{\eta ^2}{16}\right)^{\frac{4+\et^2}{4-\eta ^2}} 
 N^2 }{ 2^{\frac{64 + (4-\et^2)^2}{8(4-\eta ^2)}  } 
\left(20-\eta ^2\right) \left(4+\eta ^2\right)^{\frac{4+\eta ^2}{4-\eta ^2}}  
\Gamma \left(\frac{3}{4}-\frac{\eta ^2}{16}\right)^{\frac{12-\eta ^2}{4-\eta ^2}} 
\kappa ^2  \ L}  .
\eea
Following the same strategy used in the previous section, 
for a fixed particle number $N_A$ the excitation
of the ground states leads to the following entanglement entropy change
\bea
\D S_E &=& \frac{
\left(4-\eta ^2\right)^2 \ \Gamma \left(\frac{1}{4}-\frac{\eta ^2}{16}\right)^3  \ l^2 L  }{
2^{\frac{7}{2}+\frac{\eta ^2}{8}}  \left(8-\eta ^2\right) \left(4+\eta ^2\right)  \ \Gamma \left(1-\frac{\eta ^2}{8}\right)  \Gamma \left(\frac{3}{4}-\frac{\eta ^2}{16}\right) \kappa ^2 }
\nn
&& \times
\ \lb  r_h^{\fr{12-\et^2}{4+\et^2}}  \ls 1 + \fr{4+\et^2}{16 \ l^2 L^2} \fr{N_A^2}{r_h^{\fr{16}{4+\et^2}}} \rs - r_0^{\fr{12-\et^2}{4+\et^2}}  \ls 1 + \fr{4+\et^2}{16 \ l^2 L^2} \fr{N_A^2}{r_0^{\fr{16}{4+\et^2}}} \rs \rb ,
\eea
where $r_0$ is the horizon in the extremal limit. This entanglement entropy in the nonconformal
medium satisfies the thermodynamics-like relation
\be
\lp \D E \right|_{N_A}= T_E  \ \lp \D S_E \right|_{N_A} ,
\ee
with the following entanglement temperature
\be
T_E =  \fr{  2^{\frac{11}{2}+\frac{\eta ^2}{8}}  \left(8-\eta ^2\right)  \ \Gamma \left(1-\frac{\eta ^2}{8}\right)  \Gamma \left(\frac{3}{4}-\frac{\eta ^2}{16}\right) }{ \left(4-\eta ^2\right)^2    
 \ \Gamma \left(\frac{1}{4}-\frac{\eta ^2}{16}\right)^3  }  \ \fr{1}{l}  .
\ee
This result shows that the universality of the entanglement temperature also appears
in the nonconformal medium. 

Now, let us turn to a entanglement chemical potential. 
The global quench corresponding to the insertion of particles leads to changes of
the entanglement entropy and energy.
Similar to the previous RNAdS black brane case, these two quantities satisfies the thermodynamics-like
relation
\be   \la{res:genglobalquench}
T_E \lp \D S_E  \right|_{T_H} = \lp \D E \right|_{T_H} - \m_E \ \D N_A ,
\ee
where the increased energy and the entanglement chemical potential are given by
\bea
\lp \D E \right|_{T_H}  &=&
\frac{ 4 \lb \left(\eta ^2+4\right)^2 N_A^2-16 \left(\eta ^2-12\right) l^2 L^2 r_h^{\frac{16}{\eta ^2+4}} \rb N_A }{
\lb \left(4+\eta ^2 \right)^2 \left(12+ \eta ^2 \right) N_A^2+16 \left(48 -16 \eta ^2
+ \eta ^4 \right) l^2 L^2 r_h^{\frac{16}{4+\eta ^2}} \rb \kappa ^2 l L r_h} 
\ \Delta N_A  \nn
\m_E &=& \frac{
2^{\frac{32 - 8 \eta ^2 -\eta ^4}{2 \left(4-\eta ^2\right)}} 
\pi ^{\frac{4-2 \eta ^2}{4-\eta ^2}}  \left(8-\eta ^2\right) 
\left(4-\eta ^2\right)^{\frac{2 \eta ^2}{4-\eta ^2}}
 \Gamma \left(1-\frac{\eta ^2}{4}\right)  N_A }{
 \left(20-\eta ^2\right)  \left(4+\eta ^2\right)^{\frac{4+\eta ^2}{4-\eta ^2}}  
\Gamma \left(\frac{1}{4}-\frac{\eta ^2}{16}\right)^{\frac{8-4 \et^2}{4-\eta ^2}} 
 \Gamma \left(\frac{3}{4}-\frac{\eta ^2}{16}\right)^{\frac{8}{4-\eta ^2}}
\kappa ^2 L  }  \  \fr{1}{l^{\frac{2 \eta ^2}{4-\eta ^2} }} .
\eea
In the nonconformal medium, the entanglement
chemical potential is proportional to the width of the strip with an appropriate power, 
$\sim l^{-\frac{2 \eta ^2}{4-\eta ^2} }$. This result shows that the size dependence of
the entanglement chemical potential 
crucially depends on the nonconformality unlike the entanglement temperature.
Since $\lp \D E \right|_{T_H}$ in \eq{res:genglobalquench} is 
more dominant than $ \m_E \ \D N_A$ in the limit with small $l$ and $\et$,
the global quench at zero temperature creates the following entanglement entropy
\be			\la{res:increenzero}
\lp \D S_E  \right|_{T_H} \sim \fr{\lp \D E  \right|_{T_H}}{T_E} \sim  l^{\frac{1}{2}+\frac{\eta ^2}{8}}  \ L^{- \frac{1}{2}+\frac{\eta ^2}{8}}  \  N_A^{\frac{1}{2}-\frac{\eta ^2}{8}} \ \D N_A .
\ee

\section{Conclusion}

We have studied the holographic entanglement entropy of a nonconformal  medium
and its thermodynamics-like relation. 
In a usual thermal system including the matter, the number of the matter 
is regarded as a fundamental variable which together with temperature and volume describes a canonical ensemble.
In this case, the additional fundamental variable changes the first law of thermodynamics. 
In the nonconformal medium dual to the charged dilaton black brane, 
the similar modification happens in the thermodynamics-like relation of the 
entanglement entropy. To describe such a modification, we need to introduce
a corresponding new variable called the entanglement chemical potential. This new quantity
describes the chemical potential caused by the correlation between quantum states.
If the entanglement temperature and entanglement chemical potential are defined properly,   
the entanglement entropy in a thin strip satisfies 
the modified thermodynamics-like relation. In addition,
the entanglement temperature in a nonconformal medium still shows a universal behavior inversely 
proportional to the size of the subsystem, while the size dependence of the 
entanglement chemical potential crucially relys on the nonconformality parameter.

In this paper, we further showed that the thermodynamics-like relation 
can describe the entanglement entropy change of ground and excited states 
under the insertion of particles. If particles are added uniformly and suddenly, 
it represents the global deformation called the global quench.
Using the regularity of the bulk gauge field, the insertion of particles can be reinterpreted 
as the change of the chemical potential in a grandcanonical ensemble. 
If adding particles at zero temperature, the global quench increases the energy 
and entanglement entropy of ground states as \eq{res:increenzero}.
It would be interesting to compare this result with data of the condensed matter system.

\vspace{1cm}

{\bf Acknowledgement}

C. Park was supported by Basic Science Research Program through the National Research Foundation of Korea(NRF) funded by the Ministry of Education (NRF-2013R1A1A2A10057490) 
and also by the National Research Foundation of Korea (NRF) grant funded by the Korea government (MSIP) (2014R1A2A1A01002306).

\vspace{1cm}

\appendix

\section{Thermodynamics of a charged black brane}

\renewcommand{\theequation}{A.\arabic{equation}}
\setcounter{equation}{0}

Assuming that there exist several roots satisfying $f(r)=0$, the largest root denoted by $r_h$ corresponds to the event horizon. If the intrinsic free parameter $\et$ is fixed, the charged black brane geometry can be classified by two black brane hairs, $m$ and $Q$. From the fact $f(r_h) = 0$, the black hole mass can be rewritten as
\be  	\la{rel:massrh}
m = r_{h}^{a} + \frac{b}{r_{h}},
\ee
so we can describe the charged black brane in terms of $r_h$ and $Q$ instead of $m$ and $Q$.
Furthermore, the Hawking temperature $T_H$ defined by the surface gravity at the horizon is given by
\be		\la{rel:temprh}
T_H = \frac{12-\eta^2}{4 \pi (4+\eta^2)}  r_h^{(4-\eta^2)/(4+\eta^2)} \ls 1 
- \frac{(4 + \et^2)^2 }{ 16 (12 - \et^2)}  \fr{Q^2}{r_h^{16 /(4 + \eta^2)}} \rs.
\ee
From this relation, $r_h$ can be implicitly reinterpreted as a function of $T_H$ and $Q$. 
As a result, we can investigate the thermodynamics of the charged black brane in terms
of the Hawking temperature and charge by using \eq{rel:massrh} and \eq{rel:temprh}.
The Bekenstein-Hawking entropy $S_{BH}$ is
\bea		\la{res:BHent}
S_{BH} 
&=& \frac{2 \pi V_2}{\k^2}  \ r_{h}^{8/(4 + \eta^2)} ,
\eea
where $V_2$ denotes the regularized volume in $(x,y)$ plane. 

Since the black brane  provide a well-defined thermodynamic system, the above charged black brane 
should satisfy the fundamental thermodynamic relation
\be
d E = T_H d S_{BH} - P dV_2 + \m dN ,
\ee
where $N = Q V_2$ is the total charge. When $N$ and $V_2$ are fixed, the internal energy $E$
from the Bekenstein-Hawking entropy becomes in terms of $r_h$
\be
E = \int d S_{BH} \ T_H = \fr{4 V_2}{\ls 4 + \eta ^2 \rs \k^2} \
r_h^{(12-\et^2)/(4+\et^2)}  \ls 1 + \fr{4+\et^2}{16} \fr{Q^2}{r_h^{16/(4+\et^2)}} \rs .
\ee
In the canonical ensemble, the free energy as a function of $T_H$, $V_2$ and $N$ is given by
\bea
F &=& E - T_H S_{BH} \nn
&=& - \frac{\ls 4 - \eta^2 \rs V_2}{2 \ls 4 +\eta^2 \rs \k^2 }  \
r_{h}^{(12-\eta^2)/(4+\eta^2)} \ls 1 - \fr{(12+\et^2) (4 + \et^2)}{16 (4-\et^2)}
\fr{Q^2}{r_h^{16 /(4 + \eta^2)}} \rs 
\eea
where $r_h$ should be a function of $T_H$, $N$ and $V_2$.
After tedious calculation, one can easily find other thermodynamic quantities like the entropy $S$, chemical potential $\m$ and pressure $P$, from the thermodynamic relation of the canonical ensemble
\bea  \la{res:canon}
S &=& - \lp \fr{\pa F}{\pa T_H} \right|_{N,V_2} = \fr{2 \pi V_2}{\k^2} r_h^{8/(4+\et^2)},
\la{res:thent} \\
\m &=& \lp \fr{\pa F}{\pa N} \right|_{T_H,V_2} = \fr{N}{2 \k^2 r_h V_2} , 
\la{res:thchpot} \\
P &=& - \lp \fr{\pa F}{\pa V_2} \right|_{T_H,N} =   
\frac{ 4 - \eta^2 }{2 \ls 4 +\eta^2 \rs \k^2 }  \
r_{h}^{(12-\eta^2)/(4+\eta^2)} \ls 1 + \fr{4 + \et^2}{16 V_2^2 }
\fr{N^2}{r_h^{16 /(4 + \eta^2)}} \rs .
\eea 
These thermodynamic quantities derived from the thermodynamic law
are consistent with the previous holographic renormalization results. For example, 
the Bekenstein-Hawking entropy in \eq{res:BHent} coincides with
the renormalized thermal entropy in \eq{res:thentgrand}.

Using the Legendre transformation, it is also possible to describe the charged black brane
as a grandcanonical ensemble. In this case, the most important thermodynamic function is
the grand potential as a function of $T_H$, $\m$ and $V_2$
\bea
\O &=& F - \m N \nn
&=& - \fr{\ls 4 - \et^2 \rs V_2}{2 \ls 4 + \et^2 \rs  \k^2} \ r_h^{(12-\et^2)/(4+\et^2)}
\ls 1 + \fr{(4+\et^2) \k^4}{4} \fr{\m^2}{r_h^{2 (4-\et^2)/(4+\et^2)}} \rs ,
\eea
where $r_h$ is a function of $T_H$ and $\m$ only. Then, other thermodynamic quantities
are from the thermodynamic law
\bea			\la{res:grandcanon1}
S &=& - \lp \fr{\pa \O}{\pa T_H} \right|_{\m,V_2} = \fr{2 \pi V_2}{\k^2} r_h^{8/(4+\et^2)},
\la{res:thentgrand1} \\
N &=& - \lp \fr{\pa \O}{\pa \m} \right|_{T_H,V_2} =  2 \k^2  V_2 r_h \m , 
\la{grand:in the ultra high energy scale} \\
P &=& \fr{\ls 4 - \et^2 \rs}{2 \ls 4 + \et^2 \rs  \k^2} \ r_h^{(12-\et^2)/(4+\et^2)}
\ls 1 + \fr{(4+\et^2) \k^4}{4} \fr{\m^2}{r_h^{2 (4-\et^2)/(4+\et^2)}} \rs  .
\eea

\end{document}